\documentclass{webofc}
\usepackage[varg]{txfonts}   
\usepackage{graphicx} 
\usepackage{url}
\usepackage{enumitem}

\begin{document}
\title{The integration of heterogeneous resources in the CMS Submission Infrastructure for the LHC Run 3 and beyond}

\author{
\firstname{Antonio} \lastname{Pérez-Calero Yzquierdo}\inst{1,2}\fnsep\thanks{\email{aperez@pic.es}} \and
\firstname{Marco} \lastname{Mascheroni}\inst{3} \and
\firstname{Edita} \lastname{Kizinevic}\inst{4} \and
\firstname{Farrukh Aftab} \lastname{Khan}\inst{5} \and
\firstname{Hyunwoo} \lastname{Kim}\inst{5} \and
\firstname{Maria} \lastname{Acosta Flechas}\inst{5} \and
\firstname{Nikos} \lastname{Tsipinakis}\inst{4} \and
\firstname{Saqib} \lastname{Haleem}\inst{6}~on behalf of the CMS Collaboration.}

\institute{
Centro de Investigaciones Energ\'eticas, Medioambientales y Tecnol\'ogicas (CIEMAT), Madrid, Spain \and
Port d'Informaci\'o Cientifica (PIC), Barcelona, Spain \and
University of California San Diego, La Jolla, CA, USA \and
European Organization for Nuclear Research (CERN), Geneva, Switzerland \and
Fermi National Accelerator Laboratory, Batavia, IL, USA \and
National Centre for Physics, Islamabad, Pakistan
         }

\abstract{%
While the computing landscape supporting LHC experiments is currently dominated by x86 processors at WLCG sites, this configuration will evolve in the coming years. LHC collaborations will be increasingly employing HPC and Cloud facilities to process the vast amounts of data expected during the LHC Run 3 and the future HL-LHC phase. These facilities often feature diverse compute resources, including alternative CPU architectures like ARM and IBM Power, as well as a variety of GPU specifications. Using these heterogeneous resources efficiently is thus essential for the LHC collaborations reaching their future scientific goals. The Submission Infrastructure (SI) is a central element in CMS Computing, enabling resource acquisition and exploitation by CMS data processing, simulation and analysis tasks. The SI must therefore be adapted to ensure access and optimal utilization of this heterogeneous compute capacity. Some steps in this evolution have been already taken, as CMS is currently using opportunistically a small pool of GPU slots provided mainly at the CMS WLCG sites. Additionally, Power9 processors have been validated for CMS production at the Marconi-100 cluster at CINECA. This note will describe the updated capabilities of the SI to continue ensuring the efficient allocation and use of computing resources by CMS, despite their increasing diversity. The next steps towards a full integration and support of heterogeneous resources according to CMS needs will also be reported.

}
\maketitle
\section{Interest on heterogeneous resources}
\label{sec:intro}
The present availability of compute power in non-CPU, non-x86 processor types is abundant and increasing in the main High Performance Computing (HPC) facilities~\cite{top500}. A significant fraction of the processing power in many of these clusters is already provided by accelerators (e.g. NVIDIA GPUs in the Leonardo cluster at CINECA), while non-x86 CPU architectures are present as well (e.g. IBM Power systems such as Summit, Sierra, Marconi-100). These diverse resource types are not only deployed at HPC facilities, but are also becoming relevant at WLCG~\cite{wlcg} computing sites, traditionally built on x86 architecture CPUs. The LHC experiments are getting prepared to execute a substantial part of their data processing and simulation tasks at HPCs, including their non-x86 component. In general, a wider diversity of computing resource types is expected to become the norm in computing for High Energy Physics (HEP) experiments, resulting in higher resource heterogeneity~\cite{roadmap}. 

The CMS collaboration~\cite{cms} has recently analyzed the required evolution of its computing model and resource needs, looking into the LHC Run 3 and specially at the HL-LHC phase. The ECom2X study (see, for example~\cite{ecom2x}) concludes that CMS should expand its pool of resources to facilities beyond those dedicated to the LHC, striving towards using HPC resources effectively. As a consequence, efforts should be directed towards enabling CMS software to support non x86\_64 CPU architectures, so that CMS computing tasks can be executed on heterogeneous resources (CPUs, GPUs, FPGAs and TPUs). Moreover, the CMS Phase-2 Computing Model update report~\cite{phase2_comp} also advises to abandon the assumption of uniformity of our compute resources, evolving our workload management (WM) and resource provisioning systems to embrace the growing resource heterogeneity. 

\section{The CMS Submission Infrastructure}
\label{sec:SI}
\begin{figure}[ht]
\begin{center}
\includegraphics[width=10cm]{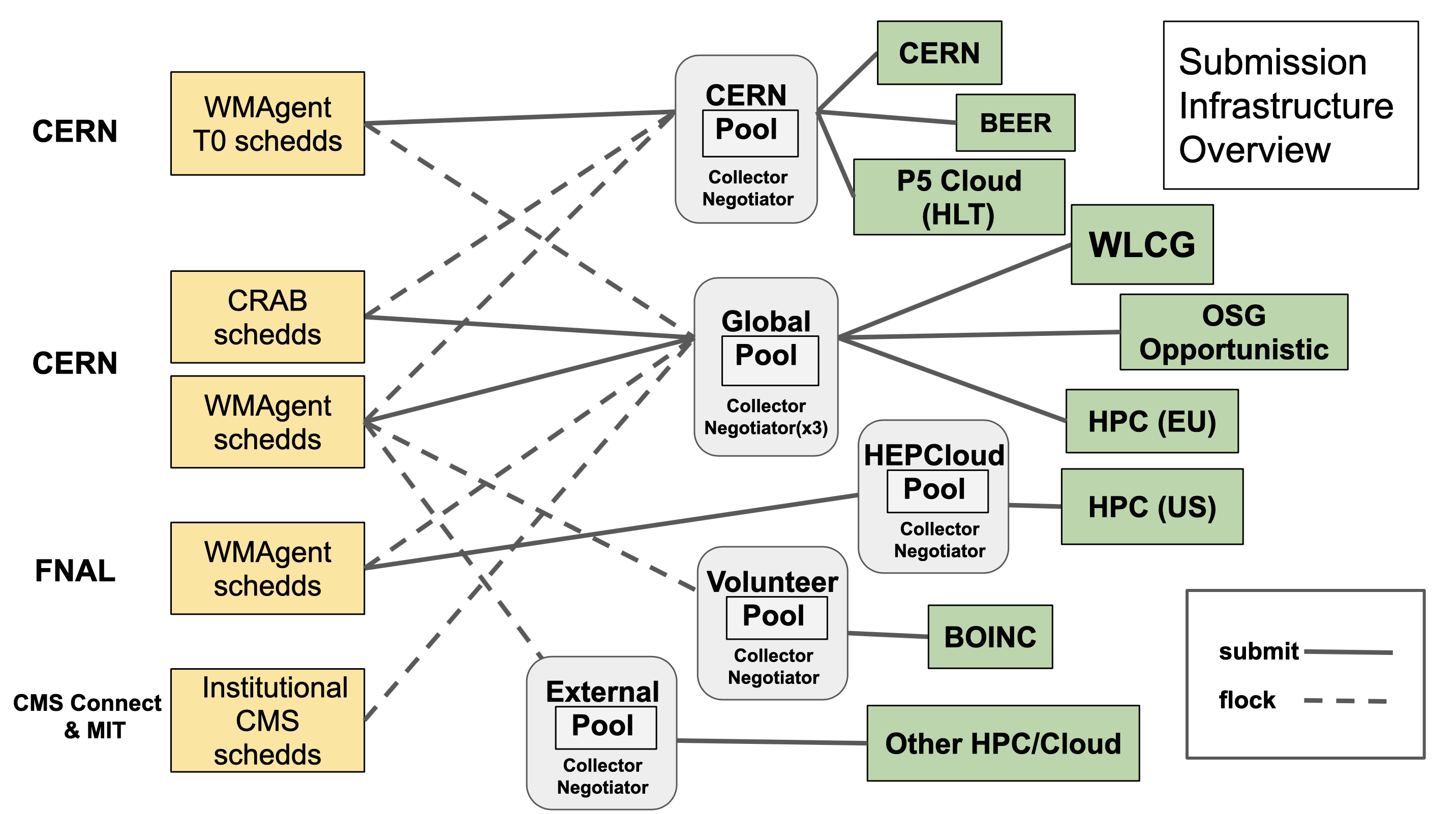}
\caption{CMS SI current configuration, including multiple federated pools of resources allocated from diverse origins (green boxes) and sets of distributed job schedulers (schedds), handling the Tier 0 and centralized production workloads (WMAgent), as well as analysis job submission (CRAB).} 
\label{fig:complexity}
\end{center}
\end{figure}

The CMS Submission Infrastructure (SI) team is in charge of operating a set of federated HTCondor~\cite{htcondor} pools, aggregating geographically distributed resources from about 70 WLCG sites, plus non-Grid resource providers, where reconstruction, simulation, and analysis of CMS physics data takes place. Figure~\ref{fig:complexity} presents a schematic view of this infrastructure. The SI is in continuous evolution (see for example~\cite{sicomplexity}), managing an ever growing collection of computing resources, connecting new and more diverse resource types and resource providers (WLCG, HPC, Cloud, volunteer). The main challenge for the SI team is to drive the evolution of our infrastructure, while maintaining the efficiency of use in all the available resources, maximizing data processing throughput and enforcing task priorities according to CMS research program. The evolution of the SI to support heterogeneous resources will be described in the following sections.

\section{Integration of GPUs to CMS SI}
\subsection{Non-pledged heterogeneous resources}
\label{sec:pledge}
As a foreword to the discussion of GPUs integration into CMS SI, it's important to note that, at the moment, all GPUs in use for CMS offline computing are opportunistic, rather than pledged resources. In contrast to pledged CPUs at WLCG sites, where agreed-upon standards exists for job execution slots (e.g. 8-core slots, 48 hours of runtime, a minimum of 2 GB per core), there is no equivalent standard job slot for the case of GPUs yet. This presents several challenges in terms of how we approach providers regarding GPU resource availability for CMS use (e.g. how do we verify the presence of GPUs at each site, what parameters are acceptable for GPU resource provisioning according to their particular policies, etc). The lack of a generic GPU slot thus deeply influences our resource provisioning and utilization strategy. Furthermore, determining the correct GPU type for each CMS task and predicting task execution parameters on these resources (including maximum execution time and memory utilization) is, at this point, a non-resolved problem. A detailed job and slot description is therefore required to ensure the appropriate slot selection, the effective use of the resources and minimizing resource wastage.

\subsection{Allocating and using resources in the CMS SI dynamic pools}
\label{sec:dynamic}
\begin{figure}
\begin{center}
\includegraphics[width=8cm]{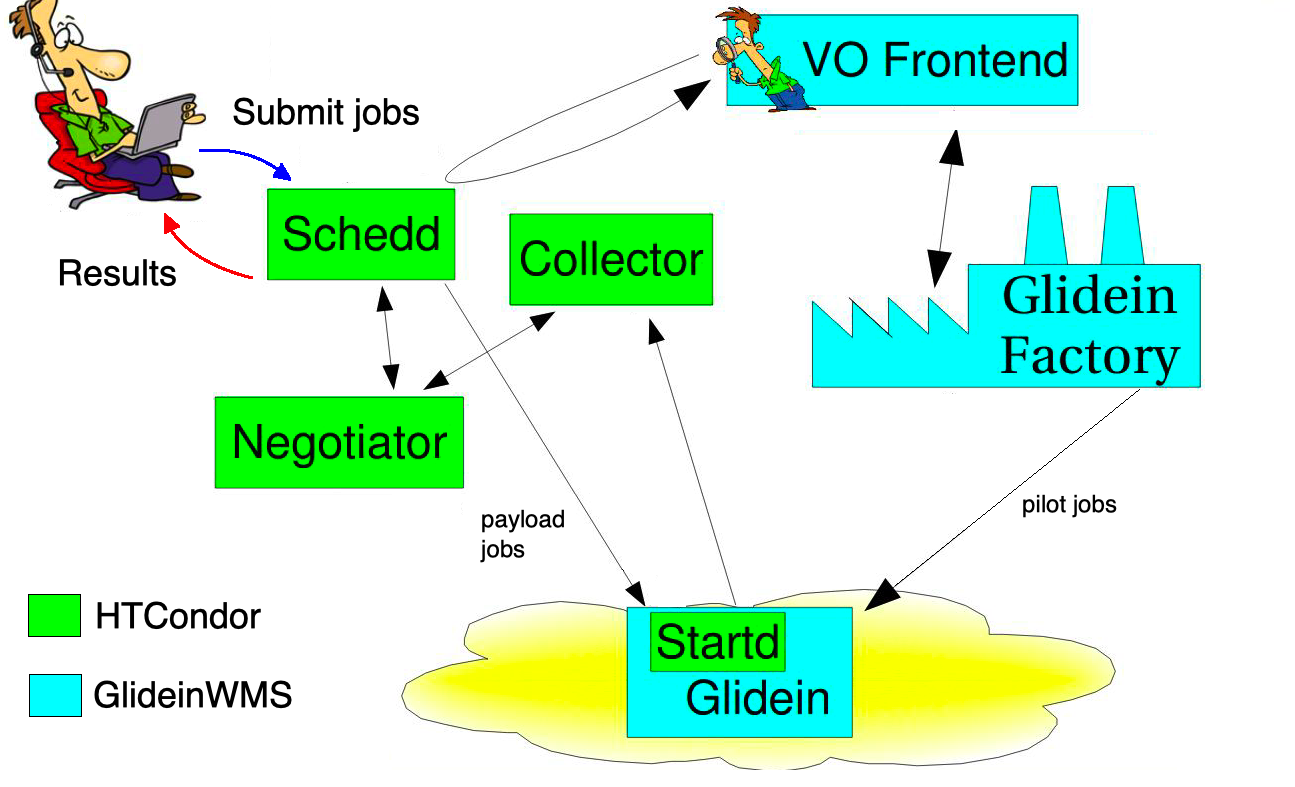}
\end{center}
\caption{GlideinWMS and HTCondor components building a dynamically sized pool of compute resources for CMS SI.}
\label{fig:pilots}
\end{figure}

The SI operates in two distinct stages of resource to requests matchmaking, corresponding first to resource allocation via GlideinWMS~\cite{gwms} pilot jobs~\cite{pilots}, followed by job to slot matchmaking managed by HTCondor. Figure~\ref{fig:pilots} highlights the main steps and components involved in each of the two phases: resource requests for each site are calculated by the Frontend (FE) component, based on user job requirements compatibility with the static description of the resources of each site held by the pilot factories. Pilot jobs are then submitted by the factories onto WLCG sites' compute elements (CE). Once requests have been accepted by the sites, allocated resources launch HTCondor startd processes that join the SI pool collector, which are then negotiated and assigned to the execution of compatible CMS tasks. Provisioning and using GPU resources just follow the same process.

\subsection{GPU resource description in GlideinWMS factories and first matchmaking}
\label{sec:factories}
The description of GPU resources available to the SI for the first matchmaking stage is encoded into pilot factory entries. Only limited and static information is available at this point, such as CE and local queues that allow access to GPU nodes, while more detailed GPU specs will only become available once the pilot jobs start. Some slot attributes are configured in agreement with the sites' policies on GPU resources allocation, such as the maximum granted task execution time. Typically, GPU slots are configured along with standard pilot features (8 CPU cores, 2 GB/core RAM). An example GPU resource entry in the pilot factory (for a HTCondor CE) looks like:
\begin{footnotesize} 
\begin{verbatim}
<entry name=" CMSHTPC_T2_US_Wisconsin_cmsgrid01_gpu" auth_method="grid_proxy" gatekeeper=
"cmsgrid01.hep.wisc.edu cmsgrid01.hep.wisc.edu:9619" gridtype="condor" ...>
<config>
(...)
   <submit_attrs>
      <submit_attr name=" +maxMemory" value=" 20000"/>
      <submit_attr name=" +xcount" value=" 8"/>
      <submit_attr name=" Request_GPUs" value=" 1"/>
   </submit_attrs>
</submit>
</config>
   <attrs>
      <attr name="GLIDEIN_CMSSite" ... type="string" value="T2_US_Wisconsin"/>
      <attr name="GLIDEIN_CPUS" ... type="string" value=" 8"/>
      <attr name="GLIDEIN_MaxMemMBs" ... type="int" value=" 20240"/>
      <attr name="GLIDEIN_Max_Walltime" ... type="int" value="216000"/>
      <attr name="GLIDEIN_Resource_Slots" ... type="string" value="GPUs,1,
      type=main"/>
      (...)
   </attrs>
</entry>
\end{verbatim}
\end{footnotesize}

At the first matchmaking stage, the FE is configured to only trigger the submission of pilot jobs for GPU nodes on account of CMS tasks requiring GPUs (note that jobs may not require, but still profit from using GPUs, see use cases described in section~\ref{sec:WM}). Job requirements are compared to pilot factory entries and pilot jobs are only submitted to matching compute elements. Not being pledged to CMS, but rather opportunistic, site preferences concerning how to use their GPUs must be respected. 

In practice this means that no CPU jobs will start on the GPU slot for a configurable time interval (e.g. the first 30 minutes of pilot runtime) to maximize the chances of matching task that do require GPUs. After this interval, either the slot can be open to matching jobs that can (but don't have to) use GPUs, or alternatively return the slot back to site, according to each site's preference. Once the GPU component of the resource is in use, our pilots try to saturate the remainder CPU part by starting smaller compatible payload jobs.

Given the late-binding aspect of our job to resource allocation, and in order to minimize wastage, once the original workload requesting GPU resources has been completely processed, any remaining resource requests (i.e. pilots queued at sites' CEs) are cancelled by the FE. This is in fact part of the resource provisioning process for CPU-only pilots as well, but it's even more critical in the case of GPUs, being opportunistic and still relatively scarce in comparison to CPUs.

\subsection{GPU resource description in HTCondor and second matchmaking}
\label{sec:matchmaking}
Once a pilot job gets access to remote resources, a number of configuration scripts are executed as part of the pilot startup phase. At this point, the HTCondor GPU discovery tool~\cite{gpudiscovery} is invoked, retrieving GPU properties relevant for matchmaking, which are updated to the slot description (the HTCondor machine classad~\cite{classad} for the slot). Additionally, a CMS custom script tests and provides a list of CUDA~\cite{cuda} supported runtimes. An example of a GPU slot classad is the following:

\begin{footnotesize}
\begin{verbatim}
CPUs = 8
TotalSlotMemory = 20000
GPUs = 2
CUDACapability = 8.0
CUDAClockMhz = 1410.0
CUDAComputeUnits = 108
CUDACoresPerCU = 64
CUDADeviceName = "NVIDIA A100-PCIE-40GB"
CUDADriverVersion = 11.3
CUDAECCEnabled = true
CUDAGlobalMemoryMB = 40536
CUDAMaxSupportedVersion = 11030
CMS_CUDA_SUPPORTED_RUNTIMES = 10.1,10.2,11.0,11.1,...
CMS_NVIDIA_DRIVER_VERSION = 515.48.07
\end{verbatim}  
\end{footnotesize}

As part of the second SI matchmaking stage, the HTCondor negotiator will compare job requirements with machine attributes for each properly configured GPU slot that joins the SI. All dynamically retrieved GPU properties can now be used by the negotiator as matchmaking attributes, allowing for a much more refined matchmaking process in comparison to that performed by the GlideinWMS FE. An example of minimal job requirements in relation to GPUs would be:
\begin{footnotesize}
\begin{verbatim}
RequestGPUs = 1
Requirements = CUDACapability >= 3 && CUDARuntime = "11.4" && GPUMemoryMB = 8000 && …
\end{verbatim}
\end{footnotesize}

Correspondingly, the matchmaking logic implemented for GPU slots is configured as certain conditions such as 
\begin{footnotesize}\begin{verbatim} Job.CUDARuntime in Machine.CMS_CUDA_SUPPORTED_RUNTIMES\end{verbatim}\end{footnotesize}.

\subsection{GPU support in the CMS WM system}
\label{sec:WM}
The CMS WM system acts as an interface between CMS workload requests and the actual submission of jobs into the HTCondor schedd queues, mapping resource requirements into concrete job attributes to be used in both matchmaking stages in the SI. In order to support workflows that require GPU resources, a new set of key/value pairs has been introduced~\cite{WMCore_gpu}. Two main attributes control how CMS tasks interact with GPUs: A first boolean tag, an attribute not part of the standard HTCondor set, has been introduced, named \verb|RequiresGPU|, which describes whether or not tasks must employ GPU slots. A second (integer) attribute, \verb|RequestGPUs|, part of the HTCondor standard set, expresses the number of GPUs to be assigned to the task in the second matchmaking stage. The combination of these two parameters in the definition of CMS workloads cover three broad use cases :
\begin{enumerate}
    \item Job must use GPUs: \verb|RequiresGPU = 1 && RequestGPUs>0|. This combination will trigger the submission of GPU pilots by the factories and produce the allocation of enough GPUs in the HTCondor matchmaking phase.
    \item Job can use GPUs: \verb|RequiresGPU = 0 && RequestGPUs>0|. This case will not produce any further GPU requests (i.e. GPU pilots will not be submitted), but still jobs can be assigned GPU slots, if available in the pool, in the second matchmaking.
    \item Job can only use CPUs: \verb|RequiresGPU = 0 && RequestGPUs=0|. This combination covers purely CPU tasks, with no GPU resources requested or matched.
\end{enumerate}

Additional attributes that further specify the requested GPU properties have been introduced (\verb|CUDACapability|, \verb|CUDARuntime| and \verb|GPUMemoryMB|). Other optional attributes can be employed as well (e.g. \verb|CUDADriverVersion|).  


\section{Availability of GPUs in CMS SI}
\label{sec:gpu_avail} 
\begin{figure}
\begin{center}
\includegraphics[width=12cm]{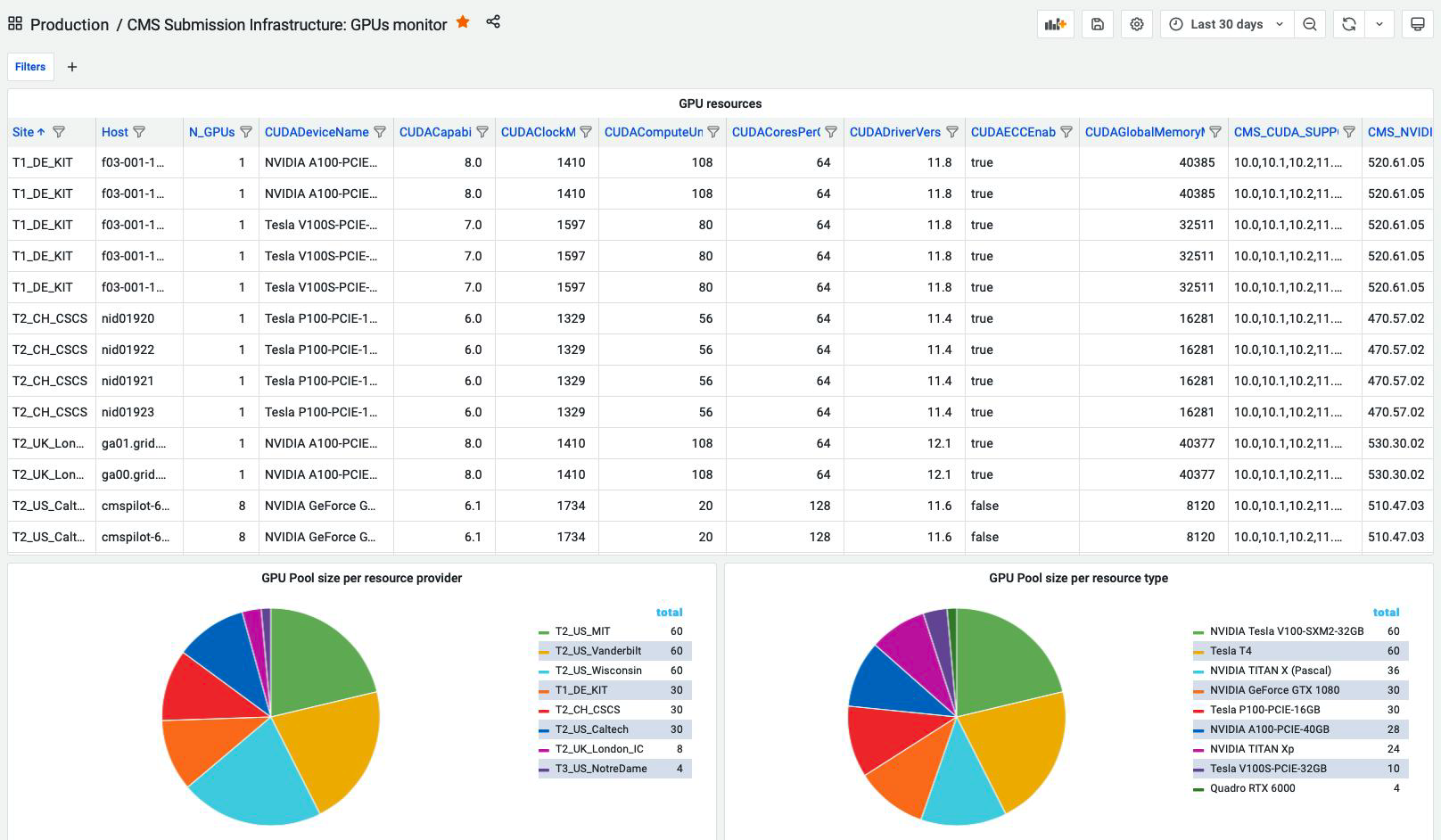}
\end{center}
\caption{GPU resources catalogue provided by the CMS SI team for CMS users, including device location and type and diverse technical properties.}
\label{fig:gpu_catalogue}
\end{figure}

Several WLCG sites are already generously contributing opportunistically their GPU resources in support of CMS computing activities. In order to provide CMS users with an updated inventory of available GPUs, the SI team conducts regular scans of the grid, gathering information on recent GPU availability, as well as their specific properties. This information is published in the CMS SI GPU monitoring table (see Figure~\ref{fig:gpu_catalogue}).

The SI team has executed a number of a scale and performance test on GPUs available to CMS. Test jobs are submitted to the SI schedds targeting any available GPUs, with the intention of matching as many as possible. Allocation of GPUs peaked at over 150 GPUs in parallel, accessing in total about 230 unique opportunistic GPUs, located mainly at CERN and US Tier-2 sites (see~\cite{gpu_acat}).

\section{Integration of non-x86 architectures} 
\label{sec:non-x86} 
The CMS SI and WM systems were originally designed and deployed with the x86\_64 standard in mind. However, both elements of CMS computing have undergone adaptations to support a broader range of processor architectures. The first non-x86 platform to be made available at scale for CMS was the Power9 architecture at CINECA's Marconi-100 cluster. In its initial phase, the extension of the SI to Power9 slots was tested using manually launched pilots executing an ad-hoc compiled HTCondor startd and without Singularity/Apptainer containers. By mid-2022, HTCondor introduced support for ARM (aarch64) and Power PC (ppc64le) CPU architectures. Startd executables for these architectures were added to the GlideinWMS pilot factory, enabling allocation and access to these new resource types through regular GlideinWMS pilots. 

The Power9 architecture has now been fully commissioned, having tested the execution of CMS tasks with valid physics results. The next focus is on ARM architecture. Both CERN and CNAF have granted access to small sets of ARM machines to be allocated for CMS into use via the SI. Integration and physics validation tests for ARM are currently ongoing. Looking ahead to the long term, there is the possibility of incorporating the RISC-V architecture, by the time of the HL-LHC era.

\begin{figure}
\begin{center}
\includegraphics[width=8cm]{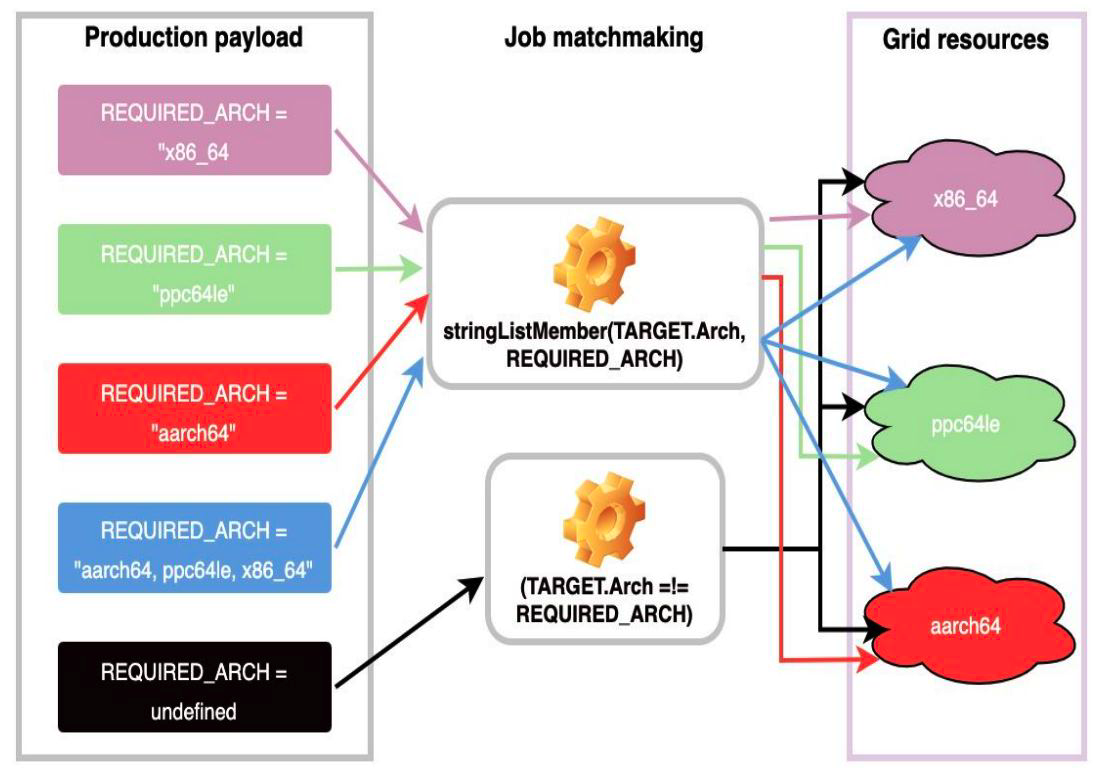}
\end{center}
\caption{Multi-architecture matchmaking in the CMS WM and SI systems.}
\label{fig:non_x86}
\end{figure}

In practice, resource architecture has become a configurable parameter in the CMS SI matchmaking processes (see Figure~\ref{fig:non_x86}). Payload jobs resource requirements may now specify a particular or multiple architectures. These resources are then provisioned and matched through the CMS SI following the general case described in Section~\ref{sec:dynamic}.

\section{Conclusions and future work}\label{sec:conclusions}
Resource heterogeneity in HEP computing is a key element moving forward. CMS is adapting its WM and SI systems to properly manage new resource types and providers. No standard slot definition exists for the emerging heterogeneous collection of resources. Workload scheduling thus requires careful description of slot properties and job requirements for an effective matchmaking and efficient use of the resources.

Regarding GPUs, they are still opportunistic but a sizable number and variety is already available for CMS users via the SI. A detailed inventory of available GPUs is regularly produced to promote GPU resource exploitation by CMS users. Non-x86 CPU architectures integration to CMS computing is ongoing, with Power9 fully integrated and validated and ARM being commissioned. 

A number of challenges towards the full exploitation of heterogeneous resources remain. These include the development of proper benchmarking and accounting, required for realistic usage at scale in the WLCG context. Another area of future development deals with the efficient execution of CMS multi-step jobs on CPU+GPU heterogeneous resources.

\section*{Acknowledgements}
This work was partially supported by the Spanish Ministry of Science and Innovation under grants PID2019-110942RB-C21, PID2019-110942RB-C22 and PID2020-113807RA-I00, which include FEDER funds from the European Union, and by the US National Science Foundation under Grant No. 2121686.


\begin{thebibliography}{}
%
\bibitem{top500} The Top500 list, \url{https://top500.org/}, accessed September, 2023.
\bibitem{wlcg} The Worldwide LHC Computing Grid \url{http://wlcg.web.cern.ch}, accessed September, 2023.
\bibitem{roadmap} The HEP Software Foundation, Albrecht, J., Alves, A.A. et al. ``A Roadmap for HEP Software and Computing R\&D for the 2020s''. Comput Softw Big Sci 3, 7 (2019). 
\bibitem{cms} CMS Collaboration, The CMS experiment at the CERN LHC, 
JINST 3 (2008) S08004, doi:10.1088/1748-0221/3/08/S08004.
\bibitem{ecom2x} T. Boccali, ``CMS Software and Offline preparation for future runs'', J. Phys.: Conf. Ser. 1525 012037 (2020).
\bibitem{phase2_comp} CMS Offline Software and Computing, ``CMS Phase-2 Computing Model: Update Document'', CMS-NOTE-2022-008 (2022).
\bibitem{htcondor} The HTCondor Software Suite public web site, \url{https://research.cs.wisc.edu/htcondor/index.html}, accessed September, 2023.
\bibitem{sicomplexity} A. Perez-Calero Yzquierdo et al. ``Evolution of the CMS Global Submission Infrastructure for the HL-LHC Era", EPJ Web Conf. 245 (2020) 03016.
\bibitem{gwms} The Glidein-based Workflow Management System, \url{https://glideinwms.fnal.gov/doc.prd/index.html}, accessed September, 2023.
\bibitem{pilots} I. Sfiligoi et al. ``The Pilot Way to Grid Resources Using glideinWMS", 2009 WRI World Congress on Computer Science and Information Engineering, Los Angeles, CA, USA, 2009, pp. 428-432, doi: 10.1109/CSIE.2009.950.
\bibitem{gpudiscovery} HTCondor gpu discovery tool, \url{https://htcondor.readthedocs.io/en/latest/man-pages/condor_gpu_discovery.html}, accessed September, 2023.
\bibitem{classad} HTCondor’s ClassAd Mechanism, \url{https://htcondor.readthedocs.io/en/latest/classads/classad-mechanism.html}, accessed September, 2023.
\bibitem{cuda} Compute Unified Device Architecture, CUDA, \url{https://developer.nvidia.com/cuda-zone}, accessed September, 2023.
\bibitem{WMCore_gpu} GPU Support in CMS WMCore, \url{https://github.com/dmwm/WMCore/wiki/GPU-Support}, accessed September, 2023.
\bibitem{gpu_acat} A. Perez-Calero Yzquierdo et al. ``Evolution of the CMS Submission Infrastructure to support heterogeneous resources in the LHC Run 3'', to be published in the proceedings of the 21st International Workshop on Advanced Computing and Analysis Techniques in Physics Research, Bari, It, 24 - 28 Oct 2022, CMS-CR-2023-035 (2023).
\end{thebibliography}
\end{document}